\documentclass[final,1p,times]{elsarticle} 
\usepackage{graphicx} 
\usepackage{amssymb} 
\usepackage{amsthm} 
\usepackage{lineno} 

\journal{Nuclear Physics A} 
\begin{document} 

\begin{frontmatter} 


\title{Looking Forward for Color Glass Condensate Signatures}

\author{Ermes Braidot$^{a}$ for the STAR collaboration}

\address[a]{Utrecht University, 
Princetonplein 5, 3584 CC, Utrecht, The Netherlands}

\begin{abstract} 
Forward $\pi^0$ production has been measured at STAR with the new Forward Meson Spectrometer (FMS) from p+p and d+Au collisions during the 2008 RHIC run. We present the first FMS results of azimuthal 
correlation involving a forward $\pi^0$ produced in p+p and d+Au collisions to search for the onset of gluon density saturation, expected to occur at small momentum fractions.
\end{abstract} 

\end{frontmatter} 



\section{Introduction}

At very high energies, scattering processes involving hadrons are dominated by
interactions of gluons, most of which carry only a small fraction
($x$) of the hadron momentum. The rapid growth of the gluon density with decreasing $x$ (accessed by higher energies) is ultimately expected to be balanced by fusion processes. These can lead eventually to a saturation of the parton density. Saturation effects are expected to be revealed in d+Au collisions (gluon densities are larger in nuclei than in nucleons) with measurements of forward hadron production, which select the small-$x$ component of the struck gluon in the nucleus. Quantifying if saturation occurs at RHIC is important because most of the matter created in heavy-ion collisions is expected to come from an initial state produced by the collisions of low-$x$ gluons in the nuclei \cite{McLerran}.

Many models try to describe forward hadron production from nuclear targets.
Saturation models \cite{saturation} include a semi-classical QCD based theory called the Color Glass Condensate (CGC) \cite{cgc,cgc2}. Other approaches include multiple scattering models \cite{mult}, shadowing models \cite{shad}, parton recombination \cite{reco}, and other descriptions that include factorization breaking in heavy nuclei \cite{fac}. The Color Glass Condensate in particular is an effective field-theory for the low-$x$ component of the hadronic wave-function in which saturation effects are associated with a new phase of the color field. Gluons with small longitudinal momenta (the relevant degrees of freedom for high-energy scattering) are described as classical gauge fields induced by a random colour source represented by the fast partons, not directly involved in the scattering. Partons form a disordered system that evolves in longitudinal momentum in a manner analogous to a glass: the dynamics of the system is given by the large-$x$ partons which propagate nearly at the speed of light, and whose internal dynamics is ``frozen'' by Lorentz time dilation \cite{what}.

\begin{figure} [ht]
\centering
\begin{center}
\includegraphics[width=0.45\textwidth, height=0.45\textwidth]{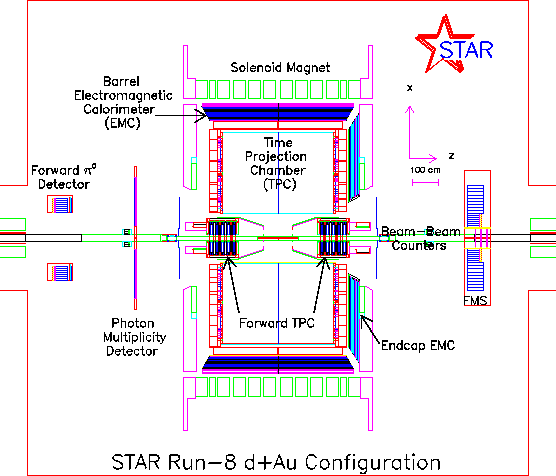}  
\includegraphics[width=0.45\textwidth]{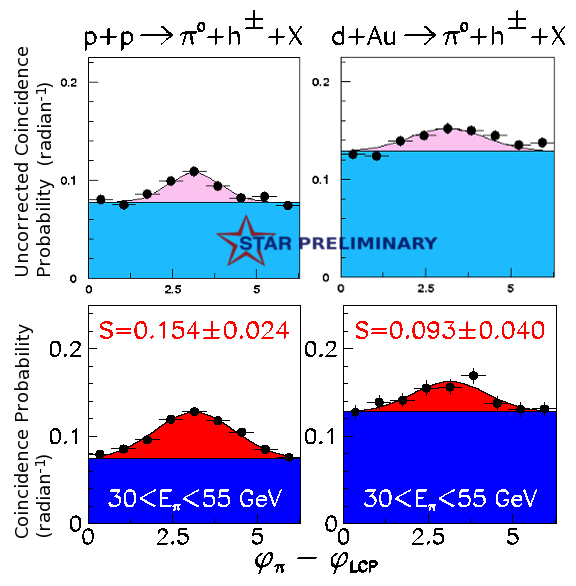} 
\caption{On the left: schematics of the STAR hall. On the right: comparison between 2003 (bottom row \cite{star1}) and uncorrected emulation from 2008 (top row) for the coincidence probability versus azimuthal angle difference between a leading forward $\pi^0$ and a leading charged particle mid-rapidity with $p_T>0.5$ GeV/c. The left (right) column is p+p (d+Au) data with statistical errors. The $\pi^0$ energy is $30<E_{\pi^0}<55$ GeV. S refers to the peak area as described in \cite{star1}.}
\label{emulation}
\end{center} 
\end{figure} 

\section{Experimental Setup}
Signatures of possible CGC effects can be searched for in pion production in the forward direction. It is in this region where high-$x$ quarks from one beam scatter from low-$x$ gluons from the other beam to produce pions in the forward direction. For this purpose forward calorimeters have been tested and improved at the STAR experiment in the first runs of the RHIC collider. Early results from the STAR experiment using prototype calorimeters 
(Forward Pion Detector FPD/FPD++) facing the deuteron beam direction of d+Au collisions, have shown \cite{star1} general agreement of inclusive yields of forward 
$\pi^0$ mesons from p+p collisions at $\sqrt{s} = 200$ GeV with NLO pQCD calculations. A stronger suppression of 
forward $\pi^0$ yield in d+Au collisions has been observed than would be expected from shadowing effects. 
Exploratory measurements of azimuthal correlations of the forward $\pi^0$ with 
charged hadrons at pseudo-rapidity $\eta\sim0$ show a recoil peak in p+p that is suppressed in 
d+Au at low pion energy. These effects are qualitatively consistent with a gluon saturation picture of the gold nucleus. A new electromagnetic calorimeter, the Forward Meson Spectrometer (FMS), has been built in place of the FPD++ and it took data for the 2008 (p+p,  d+Au) RHIC run. The FMS probes $x$ down to $x\approx10^{-4}$ \cite{GSV,lowx} for inclusive particle production at $\eta\sim4$ at $\sqrt{s_{NN}}=200$ GeV, well into the range where saturation effects are expected to set in.

\section{Analysis and Systematics}
The measurement of the azimuthal angular correlation ($\Delta\varphi$) of a forward $\pi^0$ with a coincident hadron gives important insight into particle production mechanisms. Assuming a collinear elastic parton scattering ($2\rightarrow 2$) as a first approximation, one expects to see a clear back-to-back peak centered at $\Delta\varphi=\pi$. When saturation occurs the picture is expected to change. Possibilities include a broadening of the $\Delta\varphi$ peak or even its disappearance (monojet) \cite{levin,marquet}. Effects of possible gluon saturation can be measured by comparing azimuthal correlations from p+p and d+Au interactions. Unlike the situation for Au+Au collisions where final-state  effects modify azimuthal correlations, differences between d+Au and p+p are expected to arise from initial-state effects. Wide electromagnetic coverage of the STAR detector and its full azimuthal acceptance allows us to perform azimuthal correlation analysis in a very large rapidity range ($-1.0<\eta<4.0$). A forward $\pi^0$ detected with the FMS can be correlated with a mid-rapidity $\pi^{0}$ (using the Barrel ElectroMagnetic Calorimeter, BEMC) or a mid-rapidity charge particle $h^\pm$ (using the STAR Time Projection Chamber TPC). Due to the FMS's wide acceptance and high granularity, correlations between two forward $\pi^0$'s (in the region where the lowest Bjorken-$x$ is probed) are also accessible. In the following, first results from correlation analysis using the FMS are shown. 

Consistency of the published results \cite{star1} from 2003 p+p and d+Au with new results from the FMS was checked. Azimuthal correlations between a trigger particle in the forward region and a mid-rapidity charged track $h^{\pm}$ detected by the TPC have been studied. The leading (in $p_T$) forward $\pi^0$ is selected within geometrical boundaries ($p_x>0$ , $3.8<\eta_{FMS}<4.1$) in order to emulate the 2003 acceptance. The leading charged particle (LCP) is detected in the mid-rapidity region using 2003 restrictions ($|\eta_{LPC}|<0.75$ , $p_T>0.5$~GeV/c). Correlations around $\Delta\varphi=0$ are not expected because of the large $\eta$ separation between the two particles. The data are fit with a constant function plus a Gaussian centered at $\Delta\varphi=\pi$. The comparison between 2003 FPD results and 2008 FMS emulation in both p+p and d+Au interactions appears qualitatively consistent (Fig. \ref{emulation}). Background levels and signal width are quantitatively reproduced in the $\pi^0$ energy range of $30<E_{\pi^0}<55$~GeV though more systematic studies (pile-up correction) and normalization (vertex efficiencies) are still needed in order to complete the quantitative comparisons between the published results and their emulation.

\begin{figure} 
\begin{center} 
\includegraphics[width=0.45\textwidth]{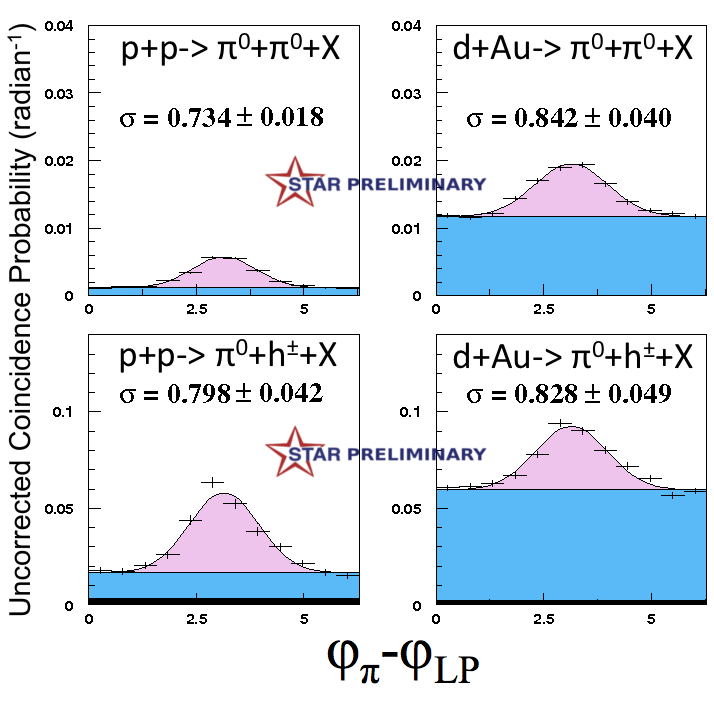} 
\includegraphics[width=0.45\textwidth]{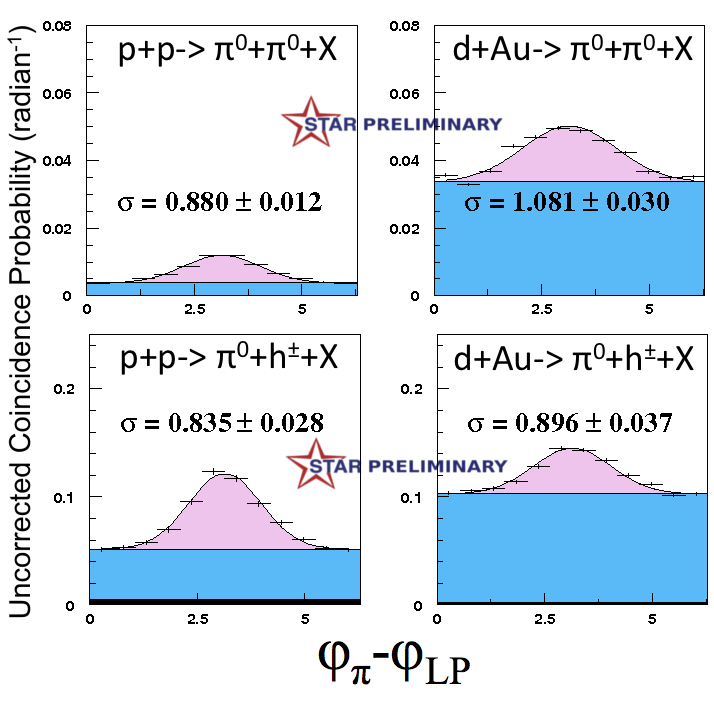} 
\caption{On the left panel: uncorrected coincidence probability versus azimuthal angle difference between a forward $\pi^0$ ($p_T^{(FMS)}>2.5$ GeV/c) and a mid-rapidity leading particle (LP, a $\pi^0$ in the upper row and a charged track $h^{\pm}$ in the bottom row) with $1.5$~GeV/c $<p_T^{(BEMC/TPC)}<p_T^{(FMS)}$. The left (right) column shows $p+p$ ($d+Au$) data with statistical errors. Systematic errors associated with luminosity dependence are now separately shown as a black band along the $x$-axis. On the right panel: same distributions with $p_T^{(FMS)}>2.0$ GeV/c , $1.0$~GeV/c $<p_T^{(BEMC/TPC)}<p_T^{(FMS)}$.}
\label{GSVcorr}
\end{center} 
\end{figure} 

The study of the azimuthal correlations can be improved by using the whole FMS acceptance and searching for associated particles in the mid-rapidity region with both TPC ($\pi^0-h^{\pm}$ correlation) and BEMC ($\pi^0-\pi^0$ correlation) detectors. A more restrictive $p_T$ cut, suggested by pQCD calculations \cite{GSV}, has been applied to this analysis. Trigger $\pi^0$'s are detected with transverse momentum $p_{T}^{(FMS)}>2.5$~GeV/c and forward pseudo-rapidity $2.5<\eta^{(FMS)}<3.5$. The associated particle is detected in the range $|\eta|<0.9$ with $1.5$~GeV/c$<p_{T}<p_{T}^{(FMS)}$. The comparison of $\Delta\varphi$ correlations shows consistency in signal width between $\pi^0-\pi^0$ and $\pi^0-h^{\pm}$ correlations (Fig. \ref{GSVcorr}, left panel) in both p+p and d+Au interactions. At this $p_T$ range the back-to-back peak at $\Delta\varphi=\pi$ is still evident in d+Au as well as in p+p. However, the comparison shows a quantitative broadening of the signal width, especially in $\pi^0_{(FMS)}-\pi^0_{(BEMC)}$ production: $\sigma_{dAu}-\sigma_{pp}=0.11\pm 0.04$. 

The broadening of the signal peak is expected to become more significant by selecting particles with lower $p_T$ (that is moving closer to the saturation scale, the boundary between possible $``$phase" regions). In this way the analysis is more sensitive to soft gluons that present a transverse size large enough to allow saturation. To this purpose a lower $p_T$ cut has been chosen for the azimuthal correlation analysis for both forward $\pi^0$ ($p_{T}^{(FMS)}>2.0$~GeV/c) and mid-rapidity particle ($1.0$~GeV/c$<p_{T}<p_{T}^{(FMS)}$). $\pi^0-\pi^0$ and $\pi^0-h^{\pm}$ azimuthal correlations for both p+p and d+Au are shown in the right panel of  Fig. \ref{GSVcorr}. As before the widths of the signal peak in $\pi^0-\pi^0$ and $\pi^0-h^{\pm}$ are consistent, both in p+p and d+Au interaction. Moreover, the broadening of the peak going from p+p to d+Au is larger than before: $\sigma_{dAu}-\sigma_{pp}=0.20\pm 0.03$. The comparison of these results using different $p_T$ cuts shows that the broadening of the signal peak is dependent on the transverse momentum of the particles that enter in the correlation.

Further systematic studies are being performed on this analysis. Azimuthal correlations have been studied also as a function of the Gold-side ($\eta<0$) multiplicity. The sum of charges has been recorded using the east side of the Beam-Beam Counter (BBC) and used as a measure of the multiplicity of the event. Luminosity dependence is studied to correct the pile-up effect for the 2008 data. In the $\pi^0-\pi^0$ correlation study an off-peak test is performed by selecting pairs of photon reconstructed in the BEMC within an invariant-mass window shifted away from the nominal $\pi^0$ peak, and simulation studies will be used to set the coincidence probability scale.

\section{Summary}

During the 2008 RHIC run, the Forward Meson Spectrometer took data and the first results look very promising. The FMS is able to reproduce the signal width of run-3 results. A comparison of $\Delta\varphi_{\pi^0_{(FMS)} +\pi^0_{(BEMC)}}$ between p+p and d+Au indicates azimuthal broadening in d+Au. Data are qualitatively consistent with a $p_T$ dependent picture of gluon saturation of the gluon nucleus.


\end{document}